\documentclass[varenna]{cimento}
\title{Composite Fermions with Spin at $\nu =1/2$.}  
\author{B. Kramer$^{1}$,
  N. Magnoli$^{2b}$, E. Mariani$^{1}$, M. Merlo$^{2a}$, F. Napoli$^{2a}$ 
 \atque M.  Sassetti$^{2a}$}
\institute{$^{1}$I. Institut f\"ur Theoretische Physik, Universit\"at Hamburg,
  Jungiusstra\ss{}e 9\\ 
20355 Hamburg, Germany\\$^{2a}$INFM-Lamia,
$^{2b}$INFN, Dipartimento di Fisica, Universit\`a di Genova, 
Via
Dodecaneso 33\\ 
16146 Genova, Italy}
      \PACSes{\PACSit{
}
{71.10.Pm, 73.43.Cd, 73.43.Nq}}
\begin{document}
\maketitle
\begin{abstract}
  The model of Composite Fermions for describing interacting electrons in two
  dimensions in the presence of a magnetic field is described. In this model,
  charged Fermions are combined with an even number of magnetic flux quanta in
  such a way that the external magnetic field is compensated on the average
  for half filling of Landau levels and the interaction is incorporated into
  an effective mass of the new composite particles. The fluctuations of the
  Chern-Simons gauge field, which describes formally the flux attachment,
  induce new interactions between the Composite Fermions. The effective
  interaction is investigated with particular emphasis on the role of the
  electron spin at filling factor $\nu=1/2$. For a system with equal numbers
  of spin-up and spin-down electrons it is found that the dominant effective
  interaction is attractive in the spin-singlet channel. This can induce a
  ground state consisting of Cooper pairs of Composite Fermions that is
  separated from the excited states by a gap. The results are used to
  understand recent spin polarization measurements done in the region of the
  Fractional Quantum Hall Effect at different constant filling factors.
\end{abstract}

\section{Introduction}

Composite Fermions are quasi-particles constructed of interacting electrons
confined to a plane and an even number of flux quanta attached to them
\cite{J89}. They have been introduced in order to explain the features in the
high-magnetic field magneto- and Hall-conductances in the region of the
Fractional Quantum Hall Effect (FQHE) \cite{TSG82}. Basically, the attachment
of fluxes is used to compensate on the average the external magnetic field at
certain (even-denominator) values of the filling factor --- the ratio between
electron density and magnetic flux density --- in such a way that the
interaction between the electrons is incorporated into single-electron
parameters as the effective mass which becomes dependent on the magnetic
field. The system of CF behaves then as a Fermi liquid of non-interacting
quasi-particles which can serve as a starting point for a perturbational
treatment \cite{HLR93}. In this model, the rational filling factors with odd
denominators at which the quantization features of the FQHE appear can be
understood as the Integer Quantum Hall Effect (IQHE) \cite{KDP80} of the
Composite Fermions \cite{H98}. A field theoretical approach to Composite
Fermions has been successfully constructed by using the path integral
formalism and by introducing a Chern-Simons gauge field \cite{LF91,KZ92}.

Direct experimental evidences for the existence of Composite Fermions have
been found by investigating the magneto-transport in spatially modulated
quantum Hall systems. The experimentally detected resistance oscillations have
been interpreted by constructing certain trajectories of quasi particles that
were commensurable with the spatial modulation of the two dimensional
electronic system \cite{W97,Setal99,WWP99,ZG99}.

Originally, the Composite Fermion description has been developed for fully
spin-polarized quantum Hall states. There are, however, experiments in the
region of low magnetic field which indicate that FQHE-ground states can be
spin unpolarized \cite{Eetal89, Eetal90}. Recently, the spin polarization of
FQHE-ground states have been measured by radiative recombination of electrons
in the inversion layer of high-electron mobility GaAs/AlGaAs-heterostructures
with holes bound to acceptors in the delta-doped region
\cite{KKE99,Ketal99,Ketal00}. The spin polarization as a function of the
magnetic field for several fixed filling factors has been investigated.
Crossovers between FQHE-ground states with different spin polarizations have
been found at certain values of the magnetic field $B$. When varying $B$, the
polarization of a ground state remains constant within a large region until a
certain crossover field $B_{\rm c}$ is reached. Then, the system is
transferred to a new differently polarized ground state which remains again
stable when $B$ is changed. 

The data have been found to be consistent with the model of non-interacting
Composite Fermions but with an effective mass that scales $\propto \sqrt{B}$.
The regions of constant spin polarization are then due to the occupation of a
fixed number of spin split Composite Fermion Landau levels. The crossover
occurs when intersections of the latter coincide with the Fermi Level. Most
strikingly, near the crossover fields, plateaus with spin polarizations almost
exactly intermediate between the fully spin polarized states appear for
temperatures extrapolated to absolute zero. This indicates additional features
beyond the non-interacting Composite Fermion model and could be signature of
partially polarized collective states. In these experiments, the system has
been tuned towards spin degeneracy by making use of the fact that the Zeeman
splitting $\Delta E_{\rm Z}$ and the cyclotron splitting $\hbar \omega_{\rm
  c}$ depend differently on $B$, $\Delta E_{\rm Z}\propto B$ and $\hbar
\omega_{\rm c}\propto \sqrt{B}$, respectively, due to the
$\sqrt{B}$-dependence of the effective mass of the Composite Fermions.

In NMR experiments \cite{Fetal01}, the spin polarization at filling factor 2/3
has also been investigated. A remarkably abrupt transition from a fully
polarized state to a state with polarization 3/4 has been detected when
decreasing the magnetic field. This has been interpreted as a first order
quantum phase transition. For filling factors higher than 2/3, a strong
depolarization has been observed that has been associated with two spin flips
per additional flux quantum. In these measurements, the system has been tuned
via tilting the magnetic field such that the Composite Fermions are subject to
an in-plane magnetic field which only influences the Zeeman splitting.

The nature of the collective states under the conditions of the FQHE including
the effect of the electron spin has been addressed in recent works. For
instance, several spin polarization instabilities have been found by assuming
the tilted field geometry \cite{GQ85,Y91}. Related to the above described
optical data, a non-translationally invariant charge density wave state of
Composite Fermions has been proposed in the basis of of restricted Hartree
Fock calculations \cite{M00}. From exact diagonalizations of a few interacting
particles, a liquid of non-symmetric excitons has been suggested for
explaining the experiments \cite{Aetal01}. It appears that these suggestions
do not exhaust the many possibilities of treating the effective interactions
between the Composite Fermions and of explaining the spin polarization
features.  Therefore, it is worthwhile to attempt different approaches. To
explain one of these is the main purpose of the present paper.

In the following chapter, we provide an introductory overview of the idea
behind the model of Composite Fermions. We show that the main feature of the
FQHE, namely the fractional quantization of the Hall conductance at fractional
filling factors, can be understood for spin polarized systems once the
fundamental idea of ''attaching flux quanta'' to a charged particle has been
accepted. In the third chapter, we consider the specific case of half filling
and explain how the formal theory of the Chern-Simons gauge transformation can
be formulated for particles including their spin degree of freedom. In the
fourth chapter, we calculate the propagators of the gauge field fluctuations
that are needed in order to understand the effective interaction between the
Composite Fermions. We find two contributions, symmetric and antisymmetric in
the gauge fluctuations, that behave differently in the small-frequency and
long-wavelength limit. In chapter five we calculate the propagators of the
Composite Fermions and establish the self-consistent equations for the
self-energies. In the sixth chapter the Dyson equation for the Composite
Fermion propagator is solved and in the seventh chapter the self-consistent
equation for the energy gap is derived.

\section{Introduction to Composite Fermion Theory}

\subsection{The Aharonov-Bohm Effect}
A qualitative understanding of the idea behind the model of Composite
Fermions can be gained by considering first the Aharonov-Bohm effect.
Consider a particle with elementary charge $e$ moving in the vector
potential ${\bf a}({\bf r})$
\begin{equation}
  \label{eq:ABvectorpotential}
 {\bf a}({\bf r})= \frac{\phi}{2\pi}\,
\frac{{\bf \hat{z}}
\times ({\bf r}-{\bf r}_{0})}{|{\bf r}-{\bf r}_{0}|^{2}}
\end{equation}
that corresponds to a localized Aharonov-Bohm magnetic field at
position ${\bf r}_{0}$ pointing into the $z$-direction (unit vector ${\bf
  \hat{z}}$)
\begin{equation}
  \label{eq:deltaflux}
  {\bf b}({\bf r})=\phi\,
\delta({\bf r}-{\bf r}_{0})\,{\bf \hat{z}}\,.
\end{equation}
The corresponding stationary Schr\"odinger equation is (light velocity $c=1$) 
\begin{equation}
  \label{eq:ABschroedinger}
  \frac{1}{2m^{*}}\,\Big[i\hbar\nabla+e{\bf a}({\bf r})\Big]^{2}
\psi({\bf r})=E\psi({\bf r})
\end{equation}
($m^{*}$ effective mass of the charge). With ${\bf r}=(x,y)$ we denote the
coordinates of the particle in the $(x,y)$-plane. By replacing (for $|{\bf
  r}-{\bf r}_{0}|\neq 0$)
\begin{equation}
  \label{eq:CSgauge}
  \psi({\bf r})=
e^{-i(e\phi/h){\rm arg}({\bf r}-{\bf r}_{0})}\,
\psi'({\bf r})\equiv 
e^{-i\tilde{\phi}\,{\rm arg}({\bf r}-{\bf r}_{0})}\,
\psi'({\bf r})
\end{equation}
with $\tilde{\phi}\equiv \phi/\phi_{0}$ ($\phi_{0}=h/e$ flux quantum), one
notes that $\psi'({\bf r})$ fulfills the Schr\"odinger equation of a free
particle,
\begin{equation}
  \label{eq:freeschroedinger}
  -\frac{\hbar^{2}\nabla^{2}}{2m^{*}}\,\psi'({\bf r})=E\psi'({\bf r})\,.
\end{equation}
In eq.~(\ref{eq:CSgauge}), ${\rm arg}({\bf r}-{\bf r}_{0})$ is the angle of
vector ${\bf r}-{\bf r}_{0}$ with the $x$-axis. Thus, eq.~(\ref{eq:CSgauge})
appears to be similar to a gauge transformation. The Aharonov-Bohm vector
potential eq.~(\ref{eq:ABvectorpotential}) is given by the gradient of the
exponent in eq.~(\ref{eq:CSgauge}). Despite the vector potential is absent in
eq.~(\ref{eq:freeschroedinger}), the wave function $\psi'({\bf r})$ contains
${\bf a}$ in the boundary conditions. Indeed, assuming in
eq.~(\ref{eq:CSgauge}) a single-valued function for $\psi({\bf r})$ implies
necessarily a multi-valued gauge-transformed function $\psi'({\bf r})$.

The above argument is completely independent of the spin of the charged
particle since the starting Hamiltonian does not couple spin and charge and
the gauge vector potential does not generate a magnetic field at the position
of the charge, ${\bf r}\neq {\bf r}_{0}$. The argument is also not changed if
an external magnetic field ${\bf B}$ is introduced. This leads to an
additional vector potential ${\bf A}({\bf r})$ and adds a spin dependent
Zeeman term $\propto {\bf \sigma}\cdot {\bf B}$ to the Hamiltonian in
eq.~(\ref{eq:freeschroedinger}) but does not influence the transformation
of eq.~(\ref{eq:CSgauge}).

\subsection{Attaching Fluxes to Fermions}

For a quantum system with $N$ charges at positions ${\bf r}_{1}\ldots{\bf
  r}_{j}\ldots{\bf r}_{N}$, interacting via a $V({\bf r}_{i}-{\bf
  r}_{j})$-potential, the above consideration may be generalized as follows.
Writing for the $N$-particle state
\begin{equation}
  \label{eq:CSstate}
  \psi_{\rm c}({\bf r}_{1}\ldots{\bf r}_{N})=\prod_{i\neq j}e^{-i\tilde{\phi}
{\rm arg}({\bf
  r}_{i}-{\bf r}_{j})}\,\psi_{\rm e}({\bf r}_{1}\ldots{\bf r}_{N})
\end{equation}
it is readily derived that if $\psi_{\rm e}$ fulfills the $N$-particle
Schr\"odinger equation in the presence of an external vector potential 
${\bf A}$ 
\begin{equation}
  \label{eq:manyparticles}
\left\{
\frac{1}{2m^{*}}\,\sum_{j}\Big[i\hbar\nabla_{j}+e{\bf A}
({\bf r}_{j})\Big]^{2}
+\frac{1}{2}\sum_{i\neq j}V({\bf r}_{i}-{\bf r}_{j})\right\}\,
\psi_{\rm e}({\bf r}_{1}\ldots{\bf r}_{N})=E\psi_{\rm e}
({\bf r}_{1}\ldots{\bf r}_{N})
\end{equation}
then $\psi_{\rm c}$ obeys the same Schr\"odinger equation, 
but with the effective vector potential
\begin{equation}
  \label{eq:vectorpotential}
  {\bf A}_{\rm eff}({\bf r}_{j})
\equiv {\bf A}({\bf r}_{j})-{\bf a}({\bf r}_{j})
\end{equation}
where 
\begin{equation}
  \label{eq:csfield}
  {\bf a}({\bf r})=\frac{\phi}{2\pi}\nabla
\sum_{i}{\rm arg}({\bf r}-{\bf r}_{i})=\frac{\phi}{2\pi}
\sum_{i}\,
\frac{{\bf \hat{z}}
\times ({\bf r}-{\bf r}_{i})}{|{\bf r}-{\bf r}_{i}|^{2}}
\end{equation}
is the generalization of eq.~(\ref{eq:ABvectorpotential}). Since $\nabla
\times {\bf a}({\bf r})=0$ for all ${\bf r}\neq {\bf r}_{j}$ ($j=1\ldots N$)
one may think of this as a gauge transformation. However, if ${\bf
  r}\to {\bf r}_{i}$ the non-single-valuedness of the phases implies
singularities at ${\bf r}_{j}$ in the gauge field,
\begin{equation}
  \label{eq:csfluxdensity}
  {\bf b}({\bf r})\equiv\nabla\times{\bf a}({\bf
  r})=\phi\,\sum_{j}\,\delta({\bf r}-{\bf r}_{j}){\bf \hat{z}}
\equiv \phi \rho({\bf r}){\bf \hat{z}}
\equiv\tilde{\phi} \phi_{0}\rho({\bf r}){\bf \hat{z}}\,,
\end{equation}
with the density of the particles
\begin{equation}
  \label{eq:density}
   \rho({\bf r})=\sum_{j}\,\delta({\bf r}-{\bf r}_{j})\,.
\end{equation}
Since the above transformation changes only the phase of the $N$-particle
wave function the probability density is not influenced 
\begin{equation}
  \label{eq:chargedensity}
  |\psi_{\rm e}({\bf r}_{1}\ldots{\bf r}_{N})|^{2}=
|\psi_{\rm c}({\bf r}_{1}\ldots{\bf r}_{N})|^{2}\,.
\end{equation}

There are several peculiarities associated with the above transformation which
--- similar as in the Aharonov-Bohm case --- ''attaches'' $\tilde{\phi}$ flux
quanta to each particle. If the number of flux quanta is an integer, the state
in eq.~(\ref{eq:CSstate}) is a multivalued function of the angles of ${\bf
  r}_{i}-{\bf r}_{j}$ with respect to rotations about multiples of $2\pi$.
When $\tilde{\phi}=2m$ ($m=1,2,3,\ldots$) the phase factors do not change the
symmetry of the $N$-particle state $\psi_{\rm c}$. If $\psi_{\rm e}$ describes
Fermions, so does $\psi_{\rm c}$, since when interchanging two particles,
${\bf r}_{i}\leftrightarrow{\bf r}_{j}$, the corresponding phase factor is
even. On the other hand, when $\tilde{\phi}=2m+1$, the phase factor changes
sign. In this case, the new state $\psi_{\rm c}$ is a Boson state. For
non-integer $\tilde{\phi}$, particles with intermediate symmetries --- anyons
--- can be generated \cite{Aetal85,IL92}.

\subsection{The Mean Field Approximation}

By introducing the mean gauge field ($\rho$ mean particle number
density)
\begin{equation}
  \label{eq:meanfield}
  \overline{{\bf b}}=\tilde{\phi}\phi_{0}\rho{\bf \hat{z}}
\end{equation}
the gauge field may be used to compensate on the average the external magnetic
field
\begin{equation}
  \label{eq:compensation}
   {\bf A}_{\rm eff}({\bf r})\equiv {\bf A}({\bf r})-
\overline{{\bf a}}({\bf r})
+\overline{{\bf a}}({\bf r})-{\bf a}({\bf r})=
\overline{{\bf a}}({\bf r})-{\bf a}({\bf r})
\equiv -\delta{\bf a}({\bf r})\,.
\end{equation}
This can be achieved if ${\bf B}({\bf r})- \overline{{\bf b}}({\bf r})=0$.
Introducing the filling factor $\nu=\rho\phi_{0}/B$ and using the mean value
eq.~(\ref{eq:meanfield}) ${\overline b}({\bf
  r})=\tilde{\phi}\phi_{0}\rho\equiv B$ implies
\begin{equation}
  \label{eq:filling}
\tilde{\phi}=\frac{1}{\nu}\,.
\end{equation}
In order to compensate on the average the external field at $\nu=1/2$, two
flux quanta have to be attached to each electron. We will consider this below
in more detail.

Exactly at the filling factor where the gauge field is adjusted for
compensating the external field, the interacting electron system in the
external magnetic field becomes a system of Composite Fermions with the
external field removed, but still in the presence of gauge field
fluctuations. The Hamiltonian is now
\begin{eqnarray}
  \label{eq:CFhamiltonian}
 H&=&\frac{1}{2m^{*}}\,
\sum_{j}\Big[i\hbar\nabla_{j}-e\delta{\bf a}({\bf r}_{j})\Big]^{2}\\
&+&\frac{1}{2\phi^{2}}\int{\rm d}{\bf r}\int{\rm d}{\bf r'}\,
\nabla\times\delta{\bf a}({\bf r})\,
V({\bf r}-{\bf r'})\,\nabla'\times\delta{\bf a}({\bf r'})\,.\nonumber
\end{eqnarray}
We have used here the relation between the fluctuations of the gauge vector
potential and the gauge magnetic field 
\begin{equation}
  \label{eq:gaugefluctuation}
  \delta{\bf b}=\nabla\times\delta{\bf a}({\bf r})=\phi\rho({\bf r})
{\bf \hat{z}}\,,
\end{equation}
in order to convert the interaction between the particles into a coupling
between the gauge field fluctuations. The constant interaction terms stemming
from the mean of the gauge field have been omitted. In the kinetic energy, a
coupling between the particles and the gauge fluctuations occurs. If one
assumes that via this effective particle-gauge field coupling the interaction
between the particles can be incorporated completely into a modification of
the effective mass, $m^{*}\to m_{\rm CF}$, which will eventually depend on the
magnetic field, we observe that the Hamiltonian is that of free,
non-interacting particles. The corresponding Fermi momentum (assuming, for the
moment, complete spin polarization) is
\begin{equation}
  \label{eq:Fermiwavenumber}
  k_{\rm F}=\sqrt{4\pi \rho}=\frac{1}{\ell_{B}\sqrt{m}}\,,
\end{equation}
using eq.~(\ref{eq:filling}) with filling factor $\nu=1/2m$ ($m$
integer) and with the electron magnetic length $\ell_{B}\equiv
\sqrt{\hbar/eB}$.

\subsection{The Fractional Quantum Hall Effect}

If the external magnetic field is close to, but does not exactly
coincide with, the one at $\nu =1/2m$, one expects then that in
analogy with the zero-external field limit the kinetic energy of the
new composite particles is completely quenched, and the spectrum
consists again of Landau levels at energies $E_{n}=\hbar\omega_{\rm
  CF}(n+1/2)$ with $n>0$. With ${\overline b}=2m\rho\phi_{0}$ one
obtains an effective mean field $B_{\rm eff}=B-{\overline b}$ with the
Composite Fermion cyclotron frequency $\omega_{\rm CF}=e|B_{\rm
  eff}|/m_{\rm CF}(B)$. In mean field approximation, $m_{\rm CF}$
would be equal to $m^*$. However, taking into account the interaction
between the particles in higher order, the CF mass is renormalized,
thus introducing a dependence on the magnetic field \cite{HLR93}. If
the effective filling factor $p$ of the CF is
\begin{equation}
p=\frac{\rho\phi_0}{B_{\rm eff}}\,,
\end{equation}
the filling factor $\nu$ of the original electrons corresponds to the integer
filling of $|p|$ Composite Fermion Landau levels
\begin{equation}
  \label{eq:cfllfillings}
  \nu=\frac{p}{1+2mp}\,.
\end{equation}
For $m=1$ (compensation of the external magnetic field at half filling) one
obtains for $p=1, 2, 3 \ldots$ the sequences $\nu=1/3, 2/5, 3/7, \ldots$ and
for $p=-1,-2,-3, \ldots$ the sequences $\nu=1, 2/3, 3/5, \ldots$ which are
consistent with the filling factors at which the FQHE is observed.

Using the gauge argument originally suggested by Laughlin for explaining the
quantization of the Hall conductance in the IQHE \cite{L81} one can obtain
also the fractional quantization of the Hall conductance. In this argument,
the current is related to adiabatically changing the total electronic energy
of a metallic loop via the change of a flux piercing the loop,
\begin{equation}
  \label{eq:gaugecurrent}
  I=\frac{\Delta E}{\Delta\phi}\,.
\end{equation}
This is obtained by considering the energy change corresponding to
transferring between the edges of the loop the number of electrons that are
associated with a flux change of $\phi_{0}$. In the $p$th Composite Fermion
Landau level, the total number of flux quanta associated with one electron is
$(2mp+1)/p=2m+1/p\equiv 1/n_{p}$; the $2m$ flux quanta are due to the
Chern-Simons gauge transformation while $1/p$ is due to the occupation of the
$p$th level. Thus, the energy change per flux quantum is $\Delta
E=n_{p}e\,U_{\rm H}$ (Hall voltage $U_{\rm H}$). This gives for the Hall
conductance at filling factor $\nu=p/(2mp+1)$ the fractionally quantized
values
\begin{equation}
  \label{eq:Hallconductance}
  G_{\rm H}=\frac{I}{U_{\rm H}}=\frac{p}{2mp+1}\,\frac{e^{2}}{h}
=\nu\,\frac{e^{2}}{h}
\end{equation}
that are observed in the Fractional Quantum Hall Effect.

\section{The Chern-Simons Transformation with Spin}

Formally, the transformation to Composite Fermions (CF) can be most
straightforwardly introduced by starting from the Lagrangian of an interacting
spin-degenerate two dimensional (2D) electron system with mean density
$\rho=\rho_{\uparrow}+\rho_{\downarrow}$ in the presence of a magnetic field
by introducing the statistical Chern-Simons gauge field (from now on we
consider $\hbar=1$) \cite{LF91},
\begin{equation}
  \label{eq:Lagrangian}
  {\cal L} ({\bf r},t) = {\cal L}_{\rm F} ({\bf r},t) 
+ {\cal L}_{{\rm I}} ({\bf r},t) 
+ {\cal L}_{{\rm CS}} ({\bf r},t)\,.
\end{equation}
The first term 
  \begin{eqnarray}
  \label{eq:noninteracting}
{\cal L}_{\rm F}({\bf r},t)&=&\sum_{s=\uparrow,\downarrow}
\psi^{\dagger}_{s}({\bf r},t)
\Big\{i\partial_{t}+\mu+e a_0^{s}({\bf r},t)\Big. \\
&&\qquad-\frac{1}{2m^{*}}\Big[i\nabla+e \Big({\bf A}({\bf r})-{\bf
a}^{s}({\bf r},t)\Big)\Big]^2 \Big\}\psi_{s}({\bf r},t)\nonumber\,,
\end{eqnarray}
corresponds to the non-interacting Fermions, chemical potential $\mu$,
and the spins $s=\pm 1/2=\uparrow,\downarrow$ in the presence of the
vector potential of the homogeneous external magnetic field, ${\bf
  B}=\nabla \times {\bf A}$, and the gauge fields $(a_{0}^{s},{\bf
  a}^{s})$ with ${\bf a}^{s}=(a_x^{s},a_y^{s})$ corresponding to spin
$s$. The Lagrangian of the interaction between the electrons
\begin{equation} 
\label{eq:Coulomb}
{\cal L}_{{\rm I}}({\bf r},t)=-{\frac{1}{2}}\sum_{s,s'=\uparrow,\downarrow} 
\int {\rm d}^2
r' \rho_{s}({\bf r},t) V({\bf r}-{\bf r}') \rho_{s'} ({\bf r}',t)\,,
\end{equation}
contains the densities of the Fermions with spin orientation $s$,
$\rho_{s}({\bf r},t)\equiv \psi^{\dagger}_{s}({\bf r},t) \psi_{s}({\bf r},t)$.

For the interaction, we assume a homogeneous isotropic potential
$V({\bf r})=V(r)=V_{\lambda}/(r^{2}+d^{2})^{\lambda/2}$ ($1<\lambda
<2$). For $\lambda =1$, and $d\to 0$, this gives the pure Coulomb
repulsion with $V_{0}=e^{2}/\varepsilon$. The Fourier transform of
this is $V(q)=2\pi e^{2}/\varepsilon q$. For intermediate $\lambda$
and $d\neq 0$, the potential decays as $r^{-\lambda}$ for large $r$,
then for small $q$ we have $V(q)\propto q^{\lambda-2}$.
For $\lambda=\to 2$, $V(q\to 0)=$ const. 

The Chern-Simons Lagrangian of the gauge field
\begin{equation}
\label{eq:CS}
{\cal L}_{{\rm CS}}({\bf r},t)=-\frac{e}{\tilde{\phi}\,\phi_{0}}
\sum_{s=\uparrow\downarrow}a_0^{s}({\bf r},t)\, {\bf
\hat{z}}\cdot\nabla\times{\bf a}^{s} ({\bf r},t)\,,
\end{equation}
is responsible for attaching $\tilde{\phi}$ flux quanta $\phi_{0}\equiv hc/e =
2\pi/e$ to each Fermion (${\bf\hat{z}}$ is the unit vector in the direction
perpendicular to the plane),
\begin{equation} 
\label{eq:constraint} 
{\bf \hat{z}}\cdot\nabla\times{\bf a}^{s}({\bf r},t)\equiv b^{s}({\bf r},t)
 =\tilde{\phi}\phi_{0}\rho_{s}({\bf r},t).
\end{equation}
This can be seen by minimizing the action with respect to $a_{0}$. We have
assumed here that the gauge term does not couple the spins. This is equivalent
to assuming that the orbital and the spin degrees of freedom are completely
decoupled. The total wave function is then constructed from local individual
spin-singlet pairs. An approach to the general case has been discussed in
\cite{LF98}.

The above total Lagrangian eq.~(\ref{eq:Lagrangian}) can be shown to describe
the same system of interacting electrons moving in a plane as without the
Chern-Simons field.

With the  Chern-Simons field, the effective magnetic field acting on
an electron with the spin $s$ is given by
\begin{equation}
  \label{eq:Beff}
  B_{\rm eff}^{s}({\bf r},t)=B-b^{s}({\bf r},t)\,.
\end{equation}
If the filling factors for the two spin directions,
$\nu_{s}=\rho_{s}\phi_{0}/B$, are equal, it is possible to compensate the
external magnetic field $B$ on the average by the gauge field when adjusting
$\nu_{s}\equiv \nu = \rho_{s}\phi_{0}/B=1/\tilde{\phi}$. This implies $k_{\rm
  F}=\sqrt{2\pi\rho}$.

To be specific, we assume in the following that $\tilde{\phi}=2$, such that
the mean gauge magnetic field cancels the external one at half filling,
$\nu=1/2$. This is consistent with the above assumption of independent spin-up
and spin-down gauge fields. In this case, the phase factor introduced by the
gauge transformation into the many-electron wave function is even when
interchanging particle indices. This means that the composite particles
consisting of one electron and the two flux quanta are Fermions.

\section{The Propagator of the Gauge Field Fluctuations.}

In deriving the propagator of the gauge fields, we use the
transverse gauge, $\nabla\cdot{\bf a}^{s}=0$. Then, the Bosonic variables
associated with the gauge field fluctuations are the {\em transverse}
components of their Fourier transforms, $a_{1}^{s}({\bf q},\omega)\equiv {\bf
  \hat{z} } \cdot\hat{\bf q}\times[{\bf a}^{s}({\bf q},\omega)-\langle{\bf
  a}^{s}({\bf q},\omega)\rangle]$.  By introducing the mean gauge field into
${\cal L}_{\rm F}$ the external field ${\bf A}$ is canceled.

The total action $S=\int d{\bf r}\,dt\,{\cal L}({\bf r},t)$ 
can be written as ($\mu=0,1$; $\nu=0,1$)
\begin{eqnarray}
\label{action}
&&\qquad S=\sum_{s} \int \frac{d{\bf k}}{(2\pi)^2}\, 
\frac{d\omega}{2\pi}\,\psi^\dagger_s({\bf k}, \omega)
\left[G^{0}_s({\bf k},\omega)\right]^{-1} \psi_s({\bf k}, \omega)\\
&&\qquad\qquad+\frac{1}{2}\sum_{\alpha\mu\nu}\int\frac{d{\bf q}}{(2\pi)^2}\,
\frac{d\Omega}{2\pi}\,a_\mu^\alpha({\bf q},\Omega) 
\left[D_{\mu\nu}^{0\alpha}({\bf q},\Omega)\right]^{-1}
a^{\alpha\dagger}_\nu({\bf q},\Omega)\nonumber\\
&&\qquad\qquad+\sum_{s\mu}\int\frac{d{\bf k}}
{(2\pi)^2}\,\frac{d\omega}{2\pi}\,
\frac{d{\bf q}}{(2\pi)^2}\,\frac{d\Omega}{2\pi}
\,\psi^\dagger_s({\bf k+q},\omega+\Omega)\psi_s({\bf k},\omega)
a_\mu^s({\bf q},\Omega)v_\mu^s({\bf k,q})\nonumber
\end{eqnarray}
Here, we have defined ($\alpha=\pm$)
\begin{equation} 
\label{eq:piumeno}
a_\mu^\alpha=\frac{1}{2}(a_\mu^\uparrow+\alpha a_\mu^\downarrow)\,, 
\end{equation} 
the free Fermion propagator ($\mu$ chemical potential) 
\begin{equation}
  \label{eq:freefermion}
\left[G^0_s({\bf k},\omega)\right]^{-1}=
\omega-\frac{k^2}{2m^*} +\mu + i\delta\,{\rm sgn}\,\omega\,,
\end{equation}
and
\begin{equation}
\label{d0}
\left[D^{0\alpha}_{\mu\nu}({\bf q},\Omega)\right]^{-1}=\left(
  \begin{array}{ccc}
 0 && \frac{2ieq}{\tilde\phi \phi_0} \\
&&\\
-\frac{2ieq}{\tilde\phi \phi_0} & 
&-\frac{e^{2}\rho}{m^{*}}
-\frac{4q^2 V(q)}{\tilde\phi^2 \phi_0^2}\,\delta_{\alpha,+}\\
  \end{array} \right)\,.
\end{equation}
In eq.~(\ref{action}), the first term is due to the free electron part
of ${\cal L}_{\rm F}$ and the second term contains the contribution of
${\cal L}_{\rm CS}$, ${\cal L}_{\rm I}$, with the electron density
replaced by the constraint eq.~(\ref{eq:constraint}) and the term
$\propto {\bf a}^{2}$ in ${\cal L}_{\rm F}$ with the electron density
replaced by the mean value.  The third term represent the interactions
between the gauge field fluctuations and the electrons due to terms
$\propto {\bf a}$ in ${\cal L}_{\rm F}$. The
term stemming from replacing the charge densities in the Coulomb
interaction by the gauge fluctuations has been incorporated in the
matrix element $[D^{0\alpha}_{11}]^{-1}$ in eq.~(\ref{d0}). The vertex
connecting two Fermions with one gauge field fluctuation operator
$a_{\mu}^{s}({\bf q},\omega)$ is
\begin{eqnarray}
v^{s}_{\mu}({\bf k},{\bf q})&=&\biggl(
\begin{array}{c}
e\\
\frac{e}{m^*}\,{\bf \hat{z}} \cdot\frac{{\bf k}\times{\bf
q}}{|{\bf q}|}
\end{array}
\biggr)\,.  
\end{eqnarray}

The above action is quadratic in the gauge field fluctuations. Thus, one could
attempt to proceed by tracing out the gauge operators in order to derive the
effective interaction between the Composite Fermions. The result would be
exact, and of the form
$$D^{0}v^{2}\psi^{\dagger}\psi^{\dagger}\psi\psi,$$
independent of the frequency. This is due to the fact that the Chern-Simons
field is purely topological and does not correspond to a Hamiltonian. A
priori, the Chern-Simons gauge field does not have a dynamics. The latter is
generated only via the coupling to the Fermions. For determining the dynamics
of the {\em interacting} Composite Fermions one has to use further
approximations. By starting from $D^{0}$, the lowest order approach does not
lead to meaningful results. It is therefore important to generate the
dynamics of the gauge field fluctuations, in order to derive the effective
interaction between the Composite Fermions.

Although we are interested in the zero-temperature limit, it is more
convenient to proceed by using the finite temperature formalism for the
propagators \cite{NO88}. The propagator of the gauge field fluctuations is
defined by ($T_{\tau}$ time ordering operator)
\begin{equation}
{\cal D}_{\mu\nu}^\alpha({\bf q},\tau)=-\langle T_{\tau}
a_\mu^\alpha({\bf q},\tau) a_\nu^{\dagger\alpha}({\bf q},0) \rangle\,.
\end{equation}
Formally, it can be obtained from a Dyson equation
\begin{equation}
  \label{eq:dyson}
  {\cal D}={\cal D}^{0}+{\cal D}^{0}\,\Pi \,{\cal D}\,.
\end{equation}
The polarization $\Pi$ contains the interactions between the electrons
and the gauge fluctuations. These are treated in lowest non-vanishing
order, i. e. the last term in eq.~(\ref{action}) enters in second
order ($\Pi\approx\Pi^0$). In the limit $|\Omega_{n}|\ll v_{\rm F}q\ll
v_{\rm F}k_{\rm F}$ the result is
\begin{equation}
  \label{eq:vertex}
\Pi^{0}({\bf q},\Omega_{n})=\left(
\begin{array}{ccc}
-\frac{e^2m^{*}}{\pi}\big(1-\frac{|\Omega_{n}|}{q v_{\rm F}} \big) && 0\\
 & & \\
0 &&\frac{e^2q^2}{12\pi m^{*}}+
\frac{2e^2|\Omega_{n}| \rho}{m^{*}q v_{\rm F}}-\frac{e^{2}\rho}{m^{*}}\\
\end{array}\right)
\end{equation}
This, together with $D^0={\cal D}^0$ (cf. eq.~(\ref{d0})), gives the
propagator in random phase approximation ($\alpha=\pm 1$) 
\begin{eqnarray}
{\cal D}^{\alpha}({\bf q},\Omega_m)&=&
\frac{1}{\zeta(q)[\gamma^{+}(q)\delta_{\alpha,+}+\gamma^{-}(q)-
\eta|\Omega_m|/q]-\beta^2 q^2}\nonumber\\
&&\nonumber\\&&\nonumber\\
&&\times\left(
\begin{array}{cccc}
\gamma^{+}(q)\delta_{\alpha,+}+\gamma^{-}(q)-\eta\frac{|\Omega_m|}{q} &&
-i\beta q \\
&&&\\
i \beta q && \zeta(q) \\
\end{array}
\right)
\end{eqnarray}\\
where $\zeta(q)=e^{2}m^{*}(1-|\Omega_{n}|/q v_{\rm
  F})/\pi$, $\beta=2e/\tilde\phi \phi_0$,
$\gamma^{+}(q)=-4q^{2}V(q)/\tilde\phi^2 \phi_0^2$,
$\gamma^{-}(q)=-q^{2}e^2/12\pi m^{*}$, $\eta=2e^2\rho/m^* v_{\rm F}$.
Note that for zero Coulomb interaction $\gamma^+=0$ the symmetric and
antisymmetric propagators are equal ${\cal D}^+={\cal D}^-$.
For small $q$ and $\Omega_{n}$, the dominant matrix
elements are
\begin{eqnarray}
\label{gaugeprop1}
{\cal D}^+_{11}({\bf q},\Omega_n)&\approx&
\frac{-q}{\alpha_+(q)q^2+\alpha_-q^3
+\eta|\Omega_n|}  \\
&&\nonumber\\
\label{gaugeprop2}
{\cal D}^-_{11}({\bf q},\Omega_n)&\approx&
\frac{-q}{\alpha_{-} q^3+\eta|\Omega_n|}
\end{eqnarray}
with $\alpha_+=4qV(q)/\tilde{\phi}^{2}\phi_0^2$ and
$\alpha_-=(e^{2}/12\pi+4\pi/\tilde{\phi}^{2}\phi_{0}^{2})/m^{*}$. For
Coulomb interaction ($\lambda=1$), $V(q)\propto 1/q$ and
$\alpha^{+}\approx$ const. In this case, the matrix element ${\cal
  D}^{-}_{11}$ is much larger than ${\cal D}^{+}_{11}$ for $q\to 0$.
On the other hand, when the interaction is screened ($\lambda=2$),
$V(q\to 0)$= const, $\alpha^{+} \propto q$, and ${\cal D}^{-}_{11}$ and
${\cal D}^{+}_{11}$ are of the same order. From now on we will focus
on the unscreened Coulomb interaction ($\lambda=1$), neglecting in eq.
(\ref{gaugeprop1}) the sub-leading term $\alpha^-$.

\section{The Propagator of the Composite Fermions.}

We calculate the Green function of the Composite Fermions by starting from the
Nambu field
\begin{equation} 
\label{eq:nambu}
{\bf \Phi}({\bf k},\tau)=\left(
\begin{array}{c}
\psi_\uparrow({\bf k},\tau) \\
\psi_\downarrow^\dagger(-{\bf k},\tau) \\
\end{array}
\right) \equiv \left(
\begin{array}{c}
\Phi_1({\bf k},\tau) \\
\Phi_2^\dagger({\bf k},\tau) \\
\end{array}
\right)\,.
\end{equation}
Using the imaginary-time definition
\begin{equation}
  \label{eq:green}
 \mbox{\boldmath ${\cal G}$}({\bf k},\tau)=-\langle T_\tau {{\bf \Phi}}({\bf
k},\tau){{\bf  \Phi}}^\dagger({\bf k},0) \rangle
\end{equation}
gives the $2\times2$-matrix
\[
{\cal G}_{ij}({\bf k},\tau)=\left(
\begin{array}{cc}
-\langle T_\tau \psi_\uparrow({\bf k},\tau)\psi_\uparrow^\dagger
({\bf k},0)\rangle  &
-\langle  T_\tau \psi_\uparrow({\bf k},\tau)\psi_\downarrow
(-{\bf k},0)\rangle \\
-\langle  T_\tau \psi_\downarrow^\dagger
(-{\bf k},\tau)\psi_\uparrow^\dagger({\bf k},0)\rangle &
-\langle  T_\tau \psi_\downarrow^\dagger(-{\bf k},\tau)
\psi_\downarrow(-{\bf k},0)\rangle \\
\end{array}
\right)
\]
The diagonal parts describe the propagation of Composite Fermions with spin up
and down. The anomalous propagators in the off-diagonals, $\propto \langle
\psi_{\uparrow}\psi_{\downarrow}\rangle$, describe the propagation of
Fermion-Fermion pairs with opposite momenta and spins. We implicitly assume
that they are different from zero. This has to be verified {\em a posteriori}
at the end of the calculation. In terms of the self energy, the
frequency-dependent Matsubara Green function is given by the Dyson equation
\begin{equation}
\label{selfenergy}
\mbox{\boldmath ${\cal G}$}^{-1}({\bf k},\omega_{n})= 
\mbox{\boldmath ${\cal G}$}_{0}^{-1}({\bf k},\omega_{n})-
\mbox{\boldmath ${\it \Sigma}$}({\bf k},\omega_{n})
\end{equation}
with the free Composite Fermion Green function
\begin{eqnarray} 
\mbox{\boldmath ${\cal G}^0$}({\bf k},\omega_n)=\left(
\begin{array}{cc}
\frac{1}{i\omega_n-(k^2/2m^{*}-\mu)} & 0 \\
0 & \frac{1}{i\omega_n+(k^2/2m^{*}-\mu)} \\
\end{array}
\right)
\equiv\left(\begin{array}{cc}
\frac{1}{i\omega_n-\xi_{k}} & 0 \\
0 & \frac{1}{i\omega_n+\xi_{k}} \\
\end{array}
\right)\,.
\end{eqnarray}

In lowest non-vanishing order, the matrix elements of the self energy are
determined by the interaction terms proportional to $v_{\nu}$ $(i=1,2; j=1,2$)
\begin{eqnarray} 
\label{eq:self1}
\mathit\Sigma_{ii}({\bf k},\omega_n)&=&-\frac{1}{\beta}\sum_{\mu\nu}
(-1)^{\nu}\int\frac{d{\bf q}}
{(2\pi)^2} \sum_{\Omega_m} \left[{\cal D}_{\mu\nu}^+({\bf q},\Omega_m) +
{\cal D}_{\mu\nu}^-({\bf q},\Omega_m)\right]\times\\
&&\qquad\qquad\qquad\qquad \times v_\mu({\bf k,q})v_\nu({\bf k,-q}) 
{\cal G}_{11}({\bf k-q},\omega_n-\Omega_m)  \nonumber \\
&&\nonumber\\
\label{eq:self2}
\mathit\Sigma_{i\not= j}({\bf k},\omega_n)&=&\frac{1}{\beta}\sum_{\mu\nu}
(-1)^{\nu}\int\frac{d{\bf q}}
{(2\pi)^2}\sum_{\Omega_m} \left[{\cal D}_{\mu\nu}^+({\bf q},\Omega_m) -
{\cal D}_{\mu\nu}^-({\bf q},\Omega_m)\right]\times \\ 
&&\qquad\qquad\qquad\qquad \times v_\mu({\bf k,q})v_\nu({\bf -k,-q}) 
{\cal G}_{12}({\bf k-q},\omega_n-\Omega_m)\,.  \nonumber
\end{eqnarray}
The self-energies can be chosen to satisfy the relations \cite{AGD63}
\begin{equation}
  \label{eq:help}
\mathit\Sigma_{21}({\bf k},\omega_n)=
\mathit\Sigma_{12}({\bf k},\omega_n) 
,\qquad
\mathit\Sigma_{22}({\bf k},\omega_n)=
-\mathit\Sigma_{11}({-\bf k},-\omega_n)
\end{equation}
which can be obtained from the definitions eq.~(\ref{eq:self1}), and
eq.~(\ref{eq:self2}) and from the definition of the Nambu Green function
eq.~(\ref{eq:nambu}). The Dyson equation eq.~(\ref{selfenergy}) together with
the above eq.~(\ref{eq:self1}) and eq.~(\ref{eq:self2}) establish a
self consistent set of equations for the Green functions.

\section{Solution of the Dyson Equation.}

In order to solve the set of equations for the self energies it is useful to
transform from the Matsubara propagators to the retarded propagators via
analytic continuation to real frequencies \cite{AGD63}. One obtains for the
self-energies
\begin{eqnarray} 
\label{eq:selfretarded1}
&&\Sigma_{11}^{\rm R}({\bf k},\epsilon)=
-\frac{1}{2\pi^2}\sum_{\mu\nu}(-1)^{\nu}\int\frac{d{\bf q}}{(2\pi)^2}
\int_{-\infty}^\infty d\omega \,d\epsilon_1 \\  
&&\qquad\qquad\qquad
\frac{{\rm Im}{\left[{D}_{\mu\nu}^{+,{\rm R}}({\bf k-q},\omega) +
{D}_{\mu\nu}^{-,{\rm R}}({\bf k-q},\omega)\right]}}{\omega +
\epsilon_1-\epsilon-i\delta}\nonumber \\
&&\qquad\qquad\qquad\times v_\mu({\bf k,k-q})\,v_\nu({\bf k,q-k})\, 
{\rm Im}{{G}}^{\rm R}_{11}({\bf q},\epsilon_1)  
\left(\tanh\frac{\epsilon_1}{2T}+
\coth\frac{\omega}{2T}\right)  \nonumber \\
&& \nonumber \\
\label{eq:selfretarded2}
&&\Sigma_{12}^{\rm R}({\bf k},\epsilon)=
\frac{1}{2\pi^2}\sum_{\mu\nu}(-1)^{\nu}\int\frac{d{\bf q}}{(2\pi)^2}
\int_{-\infty}^\infty d\omega\, d\epsilon_1 \\  
&&\qquad\qquad\qquad 
\,\frac{{\rm Im}{\left[{D}_{\mu\nu}^{+,{\rm R}}({\bf k-q},\omega) -
{D}_{\mu\nu}^{-,{\rm R}}({\bf k-q},\omega)\right]}}{\omega +
\epsilon_1-\epsilon-i\delta}\nonumber \\
&&\qquad\qquad\qquad\times v_\mu({\bf k,k-q})\,v_\nu({\bf -k,q-k})\,
{\rm Im}{{G}}_{12}^{\rm R}({\bf q},\epsilon_1) 
\left( \tanh\frac{\epsilon_1}{2T}+
\coth\frac{\omega}{2T}\right)\,.\nonumber 
\end{eqnarray}

The imaginary parts of $G_{11}^R$ and $G_{12}^R$ are obtained from the
analytic continuation of ${\cal G}_{11}^R$ and ${\cal G}_{12}^R$ by observing
that the $\Sigma$-functions depend only on the modulus of the momentum in an
isotropic system
\begin{eqnarray}
G_{11}^R({\bf q},\epsilon_1)&=&\frac{\epsilon_1+\xi_q+
{\Sigma_{11}^R}^*(q,-\epsilon_1)}{[\epsilon_1-\xi_q-
\Sigma_{11}^R(q,\epsilon_1)]
[\epsilon_1+\xi_q+{\Sigma_{11}^R}^*(q,-\epsilon_1)]-
[\Sigma_{12}^R(q,\epsilon_1)]^2} \\
&& \nonumber \\
G_{12}^R({\bf q},\epsilon_1)&=&
\frac{{\Sigma_{12}^R}(q,\epsilon_1)}{[\epsilon_1-\xi_q-
\Sigma_{11}^R(q,\epsilon_1)]
[\epsilon_1+\xi_q+{\Sigma_{11}^R}^*(q,-\epsilon_1)]-
[\Sigma_{12}^R(q,\epsilon_1)]^2}
\end{eqnarray}
\begin{equation}
  \label{eq:xi}
  \xi_{q}\equiv q^{2}/2m^{*}-\mu\,.
\end{equation}
This can be rewritten in the form 
\begin{eqnarray}
G_{11}^R({ q},\epsilon_1)&=&
\frac{\epsilon_1+\xi_q-{\Sigma_{11}^R(q,\epsilon_1)}}
{[\epsilon_1-\Sigma_{11}^R(q,\epsilon_1)]^2-\xi_q^2-
[\Sigma_{12}^R(q,\epsilon_1)]^2} \\
&& \nonumber \\
G_{12}^R({q},\epsilon_1)&=&\frac{{\Sigma_{12}^R}(q,\epsilon_1)}
{[\epsilon_1-\Sigma_{11}^R(q,\epsilon_1)]^2-\xi_q^2-
[\Sigma_{12}^R(q,\epsilon_1)]^2} 
\end{eqnarray}
due to the fact that ${\rm Im}\Sigma_{11}^R$ is an even function of
$\epsilon_1$. We are interested only in the odd part of ${\rm
  Re}\Sigma_{11}^R$ since the even part only gives a correction to the
chemical potential that does not depend on the temperature \cite{E61}.

We evaluate the imaginary parts of $G_{11}^R$ and $G_{12}^R$ for small
imaginary parts of the self-energy, i.e. in the limit \mbox{${\rm
    Im}\Sigma_{11}^R,{\rm Im}\Sigma_{12}^R \to 0$}. Since we are interested in
the region of momenta next to the Fermi surface, we assume
\begin{eqnarray}
  \label{eq:Fermisurface1}
\Sigma_{11}^R(k_{\rm F},\epsilon_1)&=&\Sigma(\epsilon_1)-i\Gamma(\epsilon_1)\\
\Sigma_{12}^R(k_{\rm F},\epsilon_1)&=&\phi(\epsilon_1)-i\Theta(\epsilon_1)  
  \label{eq:Fermisurface2}
\end{eqnarray}
with $\Theta,\Gamma > 0$ because of the analytical properties of the
retarded Green functions,
\begin{eqnarray} 
 {\rm Im}G_{11}^R(q,\epsilon_1)&=&
(A+\xi_q)\,\frac{-2\Gamma A}{B^2+4\Gamma^2 A^2}  \\
 {\rm Im}G_{12}^R(q,\epsilon_1)&=&\phi\,\frac{-2\Gamma A}{B^2+4\Gamma^2 A^2}
\end{eqnarray}
with the definitions $A\equiv\epsilon_1-\Sigma(\epsilon_1)$ and
$B=A^2-\phi^2(\epsilon_1)-\xi_q^2$. For $\Gamma \to 0$ and $\Theta \to 0$ 
we get
\begin{eqnarray} 
\label{eq:img} 
{\rm Im}G_{11}^R(q,\epsilon_1)&=&-\pi\,(A+\xi_q)\,\delta(B)\,{\rm sgn}{A}\\
&=&-\pi\,{\rm sgn}{[\epsilon_1-\Sigma(\epsilon_1)]}\,
\frac{\epsilon_1-\Sigma(\epsilon_1)+\xi_q}
{2\Omega_1(\epsilon_1)}\nonumber \\
&&\qquad\times\left\{\delta[\xi_q-\Omega_1(\epsilon_1)]
+\delta[\xi_q+\Omega_1(\epsilon_1)]\right\}\nonumber \\
&&\nonumber \\
{\rm Im}G_{12}^R(q,\epsilon_1)&=&-\pi\,\phi\,\delta(B)\,{\rm sgn}{A}\\
&=&-\pi\,{\rm sgn}{[\epsilon_1-\Sigma(\epsilon_1)]}\,\frac{\phi(\epsilon_1)}
{2\Omega_1(\epsilon_1)}\nonumber \\
&&\qquad\times\left\{\delta[\xi_q-\Omega_1(\epsilon_1)]
+\delta[\xi_q+\Omega_1(\epsilon_1)]\right\}\nonumber 
\end{eqnarray} 
with 
\begin{equation}
\label{eq:omega}
\Omega_1(\epsilon_1)=
\sqrt{[\epsilon_1-\Sigma(\epsilon_1)]^2-\phi^2(\epsilon_1)}\,.
\end{equation}

In order to perform the ${\bf q}$-integrations in eq.~(\ref{eq:selfretarded1})
and eq.~(\ref{eq:selfretarded2}), we consider the dominant contribution
$D_{11}$ and rewrite the expressions for the vertices with $p\equiv|{\bf
  k-q}|$
\begin{equation}
\label{eq:vertices}
v_1({\bf k,k-q})\,v_1({\bf k,q-k})=
-\frac{e^2}{m^{*2}}\frac{k^2q^2}{p^2}\sin^2\theta
\end{equation}
where $\theta$ is the angle between ${\bf k}$ and ${\bf q}$. Aligning the
$q_x$ axis parallel the $\hat{k}$-direction, the measure is changed to
\begin{equation}
\label{eq:measure}
\int_0^\infty q{\rm d}q \int_0^{2\pi}{\rm d}\theta = 
2\int_0^\infty{\rm d}q \int_{|{k-q}|}^{k+q}{\rm d}p\,\frac{p}{k\sin\theta}
\end{equation}
with
\begin{equation}
  \label{eq:help1}
\sin\theta= \sqrt{1-\Big[\frac{k^2+q^2-p^2}{2kq}\Big]^2}\,.
\end{equation}

If we assume for the external momentum $k\approx k_{\rm F}$ and consider only
the dominant contributions due to $q\sim k_{\rm F}$ we get for
$\Sigma_{11}^{\rm R}(k_{\rm F},\epsilon)\approx\Sigma(\epsilon)$ and
$\Sigma_{12}^{\rm R}(k_{\rm F},\epsilon)\approx\phi(\epsilon)$
\begin{eqnarray}
\Sigma(\epsilon)&=&
\frac{-1}{4\pi^4}\frac{k_{\rm F}^2 e^2}{m^{*2}}\int_0^\infty{\rm d}q 
\int_0^{2k_{\rm F}}{\rm d}p\,
\sqrt{1-\frac{p^2}{4k_{\rm F}^2}}
\int{\rm d}\omega\,{\rm d}\epsilon_1\\   
&&\frac{{\rm Im}\left[{D}_{11}^{+,{\rm R}}({p},\omega) +
{D}_{11}^{-,{\rm R}}({p},\omega)\right]}{\omega +
\epsilon_1-\epsilon-i\delta}  
\,{\rm Im}{G}^{\rm R}_{11}({q},\epsilon_1) 
 \left(\tanh\frac{\epsilon_1}{2T}+
\coth\frac{\omega}{2T}\right)\nonumber\\
&&\nonumber\\
\phi(\epsilon)&=&\frac{-1}{4\pi^4}\frac{k_{\rm F}^2 e^2}{m^{*2}} 
\int_0^\infty{\rm d}q 
\int_0^{2k_{\rm F}}{\rm d}p\,\sqrt{1-\frac{p^2}{4k_{\rm F}^2}}
\int{\rm d}\omega\,{\rm d}\epsilon_1\\
&&  
\frac{{\rm Im}\left[{D}_{11}^{+,{\rm R}}({p},\omega) -
{D}_{11}^{-,{\rm R}}({p},\omega)\right]}{\omega +
\epsilon_1-\epsilon-i\delta}
\,{\rm Im}{G}_{12}^{\rm R}({q},\epsilon_1) 
 \left(\tanh\frac{\epsilon_1}{2T}+
\coth\frac{\omega}{2T}\right) \nonumber
\end{eqnarray}
since the ${D}^{\pm}_{11}$ depend only on the modulus of their argument.

The $q$-integral involves only ${\rm Im}G$ and yields, when
linearizing $\xi_q \sim v_{\rm F}(q-k_{\rm F})$,
\begin{eqnarray} 
\label{eq:intimg1}
 \int{\rm d}q\,{\rm Im}{{G}}^{\rm R}_{11}({q},\epsilon_1) 
&=& -\frac{\pi}{v_{\rm F}} {\rm sgn}{[\epsilon_1-\Sigma(\epsilon_1)]} 
\,\frac{\epsilon_1-\Sigma(\epsilon_1)}{\Omega_1(\epsilon_1)} \\
&&\nonumber\\
\label{eq:intimg2}
\int{\rm d}q\,{\rm Im}{{G}}^{\rm R}_{12}({q},\epsilon_1) 
&=& -\frac{\pi}{v_{\rm F}} {\rm sgn}{[\epsilon_1-\Sigma(\epsilon_1)]}
\,\frac{\phi(\epsilon_1)}{\Omega_1(\epsilon_1)} 
\end{eqnarray}
In evaluating this integral, some assumptions have been made. First of all,
$\Sigma_{11}$ and $\Sigma_{12}$ (cf. eqs.~(\ref{eq:Fermisurface1})),
(\ref{eq:Fermisurface2})), are assumed to have imaginary parts that are much
smaller than the real parts. We have also neglected contributions to the
self-energy that do not depend on the frequency. Although at this stage these
assumption cannot really be justified they are {\em a posteriori} found to be
consistent with the results. In any case, they are necessary in order to be
consistent with the Fermi liquid picture for the Composite Fermions. The
above equations are valid if $\Omega_1$ is real, i.e.  for
$(\epsilon_1-\Sigma)^2-\phi^2 \geq 0$, otherwise, the integrals are zero due
to the $\delta$-functions.

By introducing the quantity
\begin{equation} 
\label{eq:aux}
 \epsilon z(\epsilon)\equiv\epsilon -\Sigma(\epsilon)
\end{equation}
and defining the gap $\Delta$,
 \begin{equation}
   \label{eq:gap}
  \Delta(\epsilon)z(\epsilon)\equiv\phi(\epsilon)\,, 
 \end{equation}
such that
\begin{equation}
  \label{eq:help2}
  \Omega_1(\epsilon_1)=|{z(\epsilon_1)}|
\sqrt{\epsilon_1^2-\Delta^2(\epsilon_1)}\,,
\end{equation}
we also can write eq.~(\ref{eq:intimg1}) and eq.~(\ref{eq:intimg2}) in the
form
\begin{eqnarray} 
\label{eq:intimg3}
\int{\rm d}q\,{\rm Im}{{G}}^{\rm R}_{11}({q},\epsilon_1) 
&=&-\frac{\pi}{v_{\rm F}}
\frac{|\epsilon_1|}{\sqrt{\epsilon_1^2-\Delta^2(\epsilon_1)}} \\
&&\nonumber\\
\label{eq:intimg4}
\int{\rm d}q\,{\rm Im}{{G}}^{\rm R}_{12}({q},\epsilon_1) 
&=&-\frac{\pi}{v_{\rm F}}
\frac{{\rm sgn}\epsilon_1 
\Delta(\epsilon_1)}{\sqrt{\epsilon_1^2-\Delta^2(\epsilon_1)}} 
\end{eqnarray}

We can now perform the $p$-integrations assuming $p\ll k_{\rm F}$, thus
retaining only the first order in the square root. Using
eq.~(\ref{gaugeprop1}) and eq.~(\ref{gaugeprop2}) and defining the integrals
\begin{eqnarray}
\label{p+}
  P^+(\omega)&=&
\int_0^\infty {\rm d}p\,{\rm Im}D_{11}^{+,R}(p,\omega)=
-\frac{\pi}{4\alpha_+}{\rm sgn}\omega\\
&&\nonumber\\
\label{p-}
P^-(\omega)&=&\int_0^\infty {\rm d}p\,{\rm Im}D_{11}^{-,R}(p,\omega)=
-\frac{\pi}{3\sqrt{3}}\frac{1}{\alpha_-^{2/3}\eta^{1/3}}\omega^{-1/3}
\end{eqnarray}
we find in the limit $T\to 0$ where
\begin{equation}
\tanh\frac{\epsilon_1}{2T}\to{\rm sgn}\epsilon_1;\qquad
\coth\frac{\omega}{2T} \to {\rm sgn}\omega
\end{equation}
the expressions for the self-energies
\begin{eqnarray} 
\label{eq:sigma}
\Sigma(\epsilon)&=& \frac{1}{4\pi^3}\frac{k_{\rm F} e^2}{m^{*}} 
\int {\rm d}\omega\,{\rm d} \epsilon_1\,
\frac{{\rm sgn}{\epsilon_1}+
{\rm sgn}\omega}{\omega+\epsilon_1-\epsilon-i\delta} \, 
[P^+(\omega)+P^-(\omega)]\\
&&\qquad\qquad\times  {\rm sgn}{\epsilon_1} 
\frac{\epsilon_1}{\sqrt{\epsilon_1^2-\Delta^2(\epsilon_1)}}\nonumber\\
&&\nonumber\\
\label{eq:phi}
\phi(\epsilon)&=&\frac{1}{4\pi^3}\frac{k_{\rm F} e^2}{m^{*}} 
\int {\rm d}\omega\,{\rm d}\epsilon_1\, 
\frac{{\rm sgn}{\epsilon_1}+{\rm sgn}\omega}
{\omega+\epsilon_1-\epsilon-i\delta} \,
[P^+(\omega)-P^-(\omega)]\\
&&\qquad\qquad\times{\rm sgn}{\epsilon_1}
\frac{\Delta(\epsilon_1)}{\sqrt{\epsilon_1^2-\Delta^2(\epsilon_1)}}\,.
\nonumber  
\end{eqnarray}

Now the energy integrations have to be performed. We begin by defining the
integrals
\begin{eqnarray}
\label{f+}
F_+(\epsilon,\epsilon_1)&=&\int{\rm d}\omega\,{\rm sgn}\omega\,
 \frac{{\rm sgn}\epsilon_1+{\rm sgn}\omega}
{\omega+\epsilon_1-\epsilon-i\delta}\\
&&\nonumber\\
   F_-(\epsilon,\epsilon_1)&=&\int{\rm d}\omega \,\omega^{-1/3}\,
\frac{{\rm sgn}\epsilon_1+{\rm sgn}\omega}
{\omega+\epsilon_1-\epsilon-i\delta}
\end{eqnarray}
that must be evaluated as principal value integrals. For $F_{-}$,
\begin{eqnarray}
\label{ref-}
{\rm Re}F_-(\epsilon,\epsilon_1)&=&-\frac{\pi}{\sqrt{3}} 
\frac{[1+3\,{\rm sgn}\epsilon_1\,
{\rm sgn}(\epsilon_1-\epsilon)]}{(\epsilon-\epsilon_1)^{1/3}}\,,\\
&&\nonumber\\
\label{imf-}
{\rm Im}F_-(\epsilon,\epsilon_1)&=&\pi\,
\frac{{\rm sgn}\epsilon_1+
{\rm sgn}(\epsilon-\epsilon_1)}
{(\epsilon-\epsilon_1)^{1/3}}\,.
\end{eqnarray}
The integral $F_+$ must be calculated by introducing a cutoff
$\Lambda_{\rm c}$,
\begin{eqnarray}
\label{ref+}
{\rm Re}F_+(\epsilon,\epsilon_1)&=&
\int_{-\Lambda_{\rm c}}^{\Lambda_{\rm c}} \frac{{\rm d}\omega}
{\omega+\epsilon_1-\epsilon}+{\rm sgn}\epsilon_1
\int_{-\Lambda_{\rm c}}^{\Lambda_{\rm c}}{\rm d}\omega\, 
\frac{{\rm sgn}\omega}{\omega+\epsilon_1-\epsilon}\,,\\
&&\nonumber\\
\label{imf+}
{\rm Im}F_+(\epsilon,\epsilon_1)&=&\pi(1-{\rm sgn}\,\epsilon_1
{\rm sgn}(\epsilon_1-\epsilon)) \,.
\end{eqnarray}
We finally find for the real part
\begin{equation}
 \label{eq:realf+}
{\rm Re}F_+(\epsilon,\epsilon_1)=\log
\frac{|\Lambda_{\rm c} +\epsilon_1-\epsilon|}
{|\Lambda_{\rm c}-\epsilon_1+\epsilon|}
+{\rm sgn}\epsilon_1 \log\frac{|\Lambda_{\rm c}^2-(\epsilon_1-\epsilon)^2|}
{(\epsilon_1-\epsilon)^2} 
\end{equation}

The physically meaningful value of the cut-off can be estimated by considering
with more detail the behavior of the integral over ${\rm
  Im}D_{11}^{\pm,R}(p,\omega)$,
\begin{eqnarray}
 \int_0^{2k_{\rm F}} {\rm d}p\,{\rm Im}D_{11}^{\pm,R}(p,\omega)&=& 
-\frac{1}{2\alpha_+} 
\left(\frac{\pi}{2}-\arctan{\frac{\eta\omega}{4k_{\rm F}^2\alpha_+}}\right) 
\end{eqnarray}
This vanishes for $\omega \to \infty$. The scale for the vanishing of the
integral can be obtained by considering the argument of the $\arctan$-function
\begin{equation}
  \label{eq:arctan}
\frac{\eta\omega}{4k_{\rm F}^2\alpha_+}= 
\frac{\omega}{e^2/\epsilon l_{B}} 
\frac{1}{2k_{\rm F}l_{B}}  
\end{equation}
where $E_{\rm C}=e^2/\epsilon l_{B}$ is the energy scale of the
Coulomb interaction and $l_{\rm B}$ is the magnetic length. Therefore,
it is reasonable to choose as the cut-off $\Lambda_{\rm c}=\Lambda
k_{\rm F}l_{B}E_{\rm C}$, where $\Lambda$ represents the numerical
value of the cut-off.

\section{The Energy Gap}

In order to find the solutions of the above non-linear Eliashberg equations
eq.~(\ref{eq:sigma}) and eq.~(\ref{eq:phi}) it is convenient to define the
constant
\begin{equation}
  \label{eq:constant}
  C=\frac{1}{4\pi^{3}}\frac{k_{\rm F} e^2}{m^{*}}\,,
\end{equation}
and
\begin{eqnarray}
  \label{eq:mplusminus}
  M^+(\epsilon,\epsilon_1)&=&\int {\rm d}\omega \,P^+(\omega)\,
 \frac{{\rm sgn}\epsilon_1+{\rm sgn}\omega}
{\omega+\epsilon_1-\epsilon-i\delta}=-
\frac{\pi}{4\alpha_+} F^+(\epsilon,\epsilon_1)\,, \\
 M^-(\epsilon,\epsilon_1)&=&\int {\rm d}\omega \,P^-(\omega) \, 
\frac{{\rm sgn}\epsilon_1+{\rm sgn}\omega}
{\omega+\epsilon_1-\epsilon-i\delta}=-
\frac{\pi}{3\sqrt{3}}\frac{1}{\alpha_-^{2/3}\eta^{1/3}} 
F^-(\epsilon,\epsilon_1)\,,
\end{eqnarray}
such that
\begin{eqnarray}
\Sigma(\epsilon)&=&C\int {\rm d}
\epsilon_1 [M^+((\epsilon,\epsilon_1)+
M^-(\epsilon,\epsilon_1)] {\rm sgn}{\epsilon_1}\, 
\frac{\epsilon_1}{\sqrt{\epsilon_1^2-\Delta^2(\epsilon_1)}}  \\
\phi(\epsilon)&=&C\int {\rm d}\epsilon_1 
[M^+((\epsilon,\epsilon_1)-M^-(\epsilon,\epsilon_1)]{\rm sgn}{\epsilon_1}\, 
\frac{\Delta(\epsilon_1)}{\sqrt{\epsilon_1^2-\Delta^2(\epsilon_1)}} 
\end{eqnarray}
which give after using the definitions eqs.~(\ref{eq:aux}), (\ref{eq:gap})
\begin{eqnarray}
  \label{eq:gapequation}
\Delta(\epsilon)&=&C\int\frac{ {\rm sgn}\epsilon_1 {\rm d}\epsilon_1}
{\sqrt{\epsilon_1^2-\Delta^2(\epsilon_1)}}\\
&&\times\left\{[M_+(\epsilon,\epsilon_1)+
M_-(\epsilon,\epsilon_1)]\frac{\epsilon_1}{\epsilon}\Delta(\epsilon)
+ [M_+(\epsilon,\epsilon_1)-
M_-(\epsilon,\epsilon_1)]\Delta(\epsilon_1) \right\}\,. \nonumber
\end{eqnarray}
If we assume that the gap is energy-independent, $\Delta(\epsilon)\approx
\Delta$, this gives finally for the gap the condition
\begin{equation}
  \label{eq:finalgap}
  I_{+}+I_{-}=1\,,
\end{equation}
with quantities $I_{\pm}$ that can be calculated by expanding with respect to
$\epsilon_{1}$ around $\epsilon=0$.  We first note that the imaginary parts of
$F_+$, $F_-$ do not give contributions to the $\epsilon_1$-integral. Then,
with $|\epsilon_1|>|\Delta|$, and assuming $\Delta>0$, we get
\begin{eqnarray}
I_-&=&\frac{16\pi^2}{27}\eta^{-1/3}\alpha_-^{-2/3} 
\int_\Delta^\infty \frac{{\rm  d}
\epsilon_1 \epsilon_1^{-1/3}}{\sqrt{\epsilon_1^2-\Delta^2}}\\ 
&=&\frac{16\pi^{5/2}}{9}\eta^{-1/3}\alpha_-^{-2/3} 
\frac{\Gamma(7/6)}{\Gamma(2/3)} \Delta^{-1/3}\,,\nonumber 
\end{eqnarray}
\begin{eqnarray}
 I_+ &=&-\frac{\pi}{\alpha_+} \int_\Delta^{\Lambda_{\rm c}} {\rm d}\epsilon_1 
\frac{1}{\sqrt{\epsilon_1^2-\Delta^2}}
\left(\log\frac{\Lambda_{\rm c}+\epsilon_1}{\epsilon_1}+
\frac{\Lambda_{\rm c}}{\Lambda_{\rm c}+\epsilon_1}\right)\\
&\approx&-\frac{\pi}{\alpha_+}\left[\frac{1}{2} 
\log^2\left(\frac{\Lambda_{\rm c}}
{\Delta}\right)+
\frac{\Lambda_{\rm c}}{\sqrt{\Lambda_{\rm c}^2-\Delta^2}}
\log\left(\frac{\Lambda_{\rm c}}{\Delta}\right)\right]\nonumber   
\end{eqnarray}
in the limit $\Lambda_{\rm c}/\Delta\gg 1$.

By replacing the expressions for $\eta$ and $\alpha_{\pm}$ and
$\Lambda_{\rm c}$ we find the final result
\begin{equation}
  \label{eq:final}
  1 = C_- \left(\frac{E_{\rm F}}{\Delta}\right)^{1/3}-
  C_+ \left[\log^2\left(\Lambda'\frac{E_{\rm F}}{\Delta}\right)+
\frac{\Lambda'E_{\rm F}}{\sqrt{(\Lambda'E_{\rm F})^2-\Delta^2}}
\log\left(\Lambda'\frac{E_{\rm F}}{\Delta}\right)\right]
\end{equation}
with $\Lambda'\equiv 2\pi\Lambda/C_+$
and the  constants

\begin{equation}
  \label{eq:constants}
C_-\approx 1.4 \qquad
C_+ =\frac{E_{\rm F}}{2\pi e^2 k_{\rm F}/\varepsilon}=
\frac{E_{\rm F}}{E_{\rm C}}\frac{1}{2\pi k_{\rm F}\ell_{B}}\,.
\end{equation}
The first term in
eq.~(\ref{eq:final}) is completely independent of the interaction and
describes the contribution due to $D^{-}$. The second term is due to $D^{+}$
and stems from the interaction between particles.

Independent of the value of the magnetic field there is {\em always} a
solution $\Delta\neq 0$ to this equation. For $E_{\rm F}$ larger than $E_{\rm
  C}$ ($C_+\gg 1$) $\Delta$ becomes vanishingly small. If $E_{\rm C}$ is much
larger than $E_{\rm F}$ ($C_+\ll 1$), the gap is nearly independent of the
Coulomb interaction.

\section{Conclusion}

The non-zero solution of eq.~(\ref{eq:final}) indicates that in a {\em single}
quantum Hall layer, when two Landau levels with opposite spins intersect at
the Fermi energy in a perpendicular magnetic field, the system becomes
unstable against formation of a spin-singlet state due to an effective
attractive coupling of Composite Fermions via the gauge field fluctuations.
The resulting condensate state is similar to the macroscopic state induced in
a superconductor by the electron-phonon coupling.  Is there any experimental
indication that this indeed might be the case?

The existence of such a spin-singlet condensate state can contribute towards
the understanding of the extra-plateaus in the spin polarization experiments
of \cite{Ketal00}. The splitting between the Landau levels of the Composite
Fermions (CFLL) with spin up and spin down behaves as $\sqrt{B}$ for small
$B$, and is proportional to $B$ for large $B$ due to the Zeeman splitting.
Spin-up and spin-down components of different CFLL can intersect. As an
example, we consider $\nu\equiv p/(2p+1)=2/5$. This corresponds to two filled
CFLL ($p=2$). We adjust the Fermi level to the energy where the spin-down
Zeeman level of the lowest CFLL becomes degenerate with the spin-up Zeeman
level of the first CFLL. For magnetic fields smaller than the one
corresponding to the point of degeneracy, $B_{\rm c}$, only the Zeeman levels
of the lowest CFLL are occupied at zero temperature. The spin polarization
vanishes, $\gamma=(\rho_{\downarrow}-\rho_{\uparrow})/
(\rho_{\downarrow}+\rho_{\uparrow})=0$. Magnetic fields above $B_{\rm c}$
yield $\gamma=1$. Exactly at $B_{\rm c}$, two half-filled CFLL can be formed
when defining the filling factor in terms of the ratio between the number of
CF and the number of ''effective'' flux quanta crossing the sample.  In
analogy to the above, one could then perform a gauge transformation leading to
''second generation'' Composite Fermions with the corresponding gauge
fluctuations mediating an effective attractive interaction. This would lead to
the formation of a condensate.

The existence of the gap at the crossing point would imply that in a region of
magnetic fields around this point, where the energy difference between the
CFLL is less then $\Delta$, the condensate remains stable. The formation of
such a state of singlet CF-pairs was then responsible for the formation of a
plateau exactly at half the distance between the neighboring plateaus in an
interval of magnetic fields near the crossover point.

The possibility of generating long-range spin-pairing correlations in a single
2D Hall sample is similar to those discussed previously for QHE double layers
\cite{FW01,S00} by tuning the density and the magnetic field to induce the
crossing between spin-up and spin-down Landau levels. It leads to interesting
speculations. For instance, consider two QHE systems in the same plane, say at
$\nu=2/5$, separated by a tunnel junction. By tuning the two densities to the
value of the point of degeneracy a ''Josephson current'' should flow. Such a
current should vanish as soon as one of the two densities was detuned.

In conclusion, we have considered two Landau levels with opposite spins tuned
to intersect at filling factor $1/2$ at the Fermi level.  By applying the
Chern-Simons gauge transformation, we have derived an effective attractive
interaction between the Composite Fermions. This yields an instability towards
a spin-singlet condensate. We have discussed several experimental
consequences. In order to observe the predicted spin-singlet state, a
close-to-zero in-plane component of the magnetic field should be necessary as
has been achieved in the spin-polarization experiments done in the region of
the FQHE. 

Our results suggest that different occupations of spin-up and spin-down Landau
levels could account for instabilities at other fractional polarizations and
that an in-plane component of the magnetic field could account for an
anisotropic spin-singlet condensate.

\section*{Acknowledgment}
This work has been supported by the European Union via the TMR and RTN
programmes (FMRX-CT98-0180, HPRN-CT2000-00144), by the Deutsche
Forschungsgemeinschaft within the Schwerpunkt ``Quanten-Hall-Effekt'' of the
Universit\"at Hamburg, and by the Italian MURST via PRIN00.


\begin{thebibliography}{99}
  
\bibitem{J89} \BY{Jain~J. K.} \IN{Phys. Rev. Lett.}{63}{1989}{199}.
  
\bibitem{TSG82}\BY{Tsui~D.~C., Stormer~H. L. \atque Gossard~A. C.} \IN{Phys.
    Rev. Lett.}{48}{1982}{1559}.
  
\bibitem{HLR93}\BY{Halperin~B. I., Lee~P. A.  \atque Read~N.}  \IN{Phys. Rev.
    B}{47}{1993}{7312}.
  
\bibitem{KDP80}\BY{von Klitzing~K., Dorda~G. \atque Pepper~M.} \IN{Phys. Rev.
    Lett.}{45}{1980}{494}.
  
\bibitem{H98} \BY{Heinonen~O. (Ed.)}  \TITLE{Composite Fermions} (World
  Scientific, Singapore) 1998.

\bibitem{Aetal85}\BY{Arovas~D. P., Schrieffer~R., Wilczek~F: \atque
    Zee~A.} \IN{Nucl. Phys. B}{251[FS13]}{1985}{117}.
  
\bibitem{IL92}\BY{Iengo~R. \atque Lechner~K.} \IN{Phys. Rep.}{213}{1992}{179}.
  
\bibitem{L81}\BY{Laughlin~R. B.} \IN{Phys. Rev. B}{23}{1981}{5632}.

\bibitem{LF91}\BY{Lopez~A. \atque Fradkin~E.}  \IN{Phys. Rev.
    B}{44}{1991}{5246}.
  
\bibitem{KZ92}\BY{Kalmeyer~V.  \atque Zhang~S.  C.} \IN{Phys. Rev.
    B}{46}{1992}{9889}.

\bibitem{W97}\BY{Willett~R. L.} \IN{Adv. Phys.}{46}{1997}{447}.
  
\bibitem{Setal99}\BY{Smet~J. H., Jobst~S., von Klitzing~K., Weiss~D.,
    Wegscheider~W. \atque Umansky~V.} \IN{Phys. Rev. Lett.}{83}{1999}{2620}.

\bibitem{WWP99}\BY{Willett~R. L., West~K. W. \atque
    Pfeiffer~L. N.} \IN{Phys. Rev. Lett.}{83}{1999}{2624}.
  
\bibitem{ZG99}\BY{Zwerschke~S. D. M. \atque Gerhardts~R. R.} \IN{Phys. Rev.
    Lett.}{83}{1999}{2628}.
 
\bibitem{Eetal89}\BY{Eisenstein~J. P., Stormer~H. L., Pfeiffer~L. N. \atque
    West~K. W.} \IN{Phys. Rev. Lett.}{62}{1989}{1540}.

\bibitem{Eetal90}\BY{Eisenstein~J. P., Stormer~H. L., Pfeiffer~L. N. \atque
    West~K. W.} \IN{Phys. Rev. B}{41}{1990}{7910}.
  
\bibitem{KKE99}\BY{Kukushkin~I. V., von Klitzing~K., \atque Eberl~K.} \IN{Phys.
    Rev. Lett.}{82}{1999}{3665}.

\bibitem{Ketal99}\BY{Kukushkin~I. V., von Klitzing~K., Levchenko~K. G. \atque
    Lozovik~Yu. E.} \IN{JETP Lett.}{70}{1999}{730}.
  
\bibitem{Ketal00}\BY{Kukushkin~I. V., Smet~J. H., von Klitzing~K., \atque
    Eberl~K.} \IN{Phys.  Rev. Lett.}{85}{2000}{3688}.
  
\bibitem{Fetal01}\BY{Freytag~N., Tokunaga~Y., Horvati\'c~M., Bertier~C.,
    Shayegan~M. \atque Levy~L. P.} \IN{Phys. Rev. Lett.}{87}{2001}{136801}.
  
\bibitem{GQ85}\BY{Giuliani~G. F. \atque Quinn~J. J.}\IN{Phys. Rev.
    B}{31}{1985}{6228}.

\bibitem{Y91}\BY{Yarlagadda~S.} \IN{Phys. Rev. B}{44}{1991}{13101}.

\bibitem{M00}\BY{Murthy~G.} \IN{Phys. Rev. Lett.}{84}{2000}{350}.
  
\bibitem{Aetal01}\BY{Apalkov~V. M., Chakraborty~T., Pietilainen~P. \atque
    Miemela~K.} \IN{Phys. Rev. Lett.}{86}{2001}{1311}.

\bibitem{LF98} \BY{Lopez~A. \atque Fradkin~E.}  \TITLE{Fermionic Chern-Simons
    field theory for the Fractional Quantum Hall Effect}, in \TITLE{Composite
    Fermions}, edited by \NAME{Heinonen~O.}  (World Scientific, Singapore)
  1998, pp.~195-253.
  
\bibitem{NO88}\BY{Negele~J. W. \atque Orland~H.}  \TITLE{Quantum many-particle
    Systems} (Addison-Wesley, Redwood City) 1988.

\bibitem{AGD63}\BY{Abrikosov~A. A., Gorkov~L. P.\atque Dzyaloshinski~I. E.}
  \TITLE{Title    Systems} (Prentice Hall) 1963.

\bibitem{E61}\BY{Eliashberg~G. M.} \IN{Sov. Phys. JETP}{12}{1961}{1000}.
  
\bibitem{FW01} \BY{Fogler~M.~M. \atque Wilczek~F.} \IN{Phys. Rev. Lett.} {86}
  {2001}{1833}.
  
\bibitem{S00} \BY{Spielman~I. B., Eisenstein~J. P., Pfeiffer~ L. N., \atque
    West~K. W.} \IN{Phys. Rev. Lett.}{84}{2000}{5808}.
  
\end{thebibliography}
\end{document}